\documentclass[prb,superscriptaddress,showpacs,showkeys,twocolumn]{revtex4-1}
\usepackage{amsfonts,amsmath,amssymb,graphicx,epstopdf,verbatim}
\usepackage[english]{babel}

\newcommand{\brac}[1]{\left(#1\right)}
\newcommand{\norm}[1]{\left|#1\right|}
\newcommand{\pd}[2]{\frac{\partial #1}{\partial #2}}
\newcommand{\vectr}[2]{ \left( \begin{array}{c}
#1 \\
#2 
\end{array} \right)}
\newcommand{\matrx}[4]{ \left( \begin{array}{cc}
#1 & #2 \\
#3 & #4 
\end{array} \right)}
\renewcommand{\Im}{\text{Im}}
\renewcommand{\vec}[1]{\mathbf{#1}}

\graphicspath{{./images/}}
\begin{document}

\title{Negative drag in nonequilibrium polariton quantum fluids}
\author{Mathias Van Regemortel}
\affiliation{TQC, Universiteit Antwerpen, Universiteitsplein 1,
B-2610 Antwerpen, Belgium} 
\author{Michiel Wouters}
\affiliation{TQC, Universiteit Antwerpen, Universiteitsplein 1,
B-2610 Antwerpen, Belgium}
\date{\today}
\begin{abstract}
The possibility of a negative drag force on a defect in nonequilibrium polariton quantum fluids is presented. We relate this phenomenon to the selective parametric amplification of the waves scattered by the defect. This leads to the prediction that mobile defects acquire a non-zero velocity with respect to the polariton nonequilibrium fluid. We derive a direct relation between the drag force and the momentum distribution of the fluid, that allows for the experimental verification of our predictions.
\end{abstract}

\maketitle

\section{Introduction}
\label{sec:intro}
Exiton-polaritons in planar microcavities arise from a strong coupling between a cavity photon mode and a quantum-well exciton. Recently, these system have been the subject of intensive study in the context of out-of-equilibrium quantum fluids \cite{iac_review}. Their interest stems, on the one hand from the fact that their properties such as fluid density and velocity can be easily manipulated by an external laser field and, on the other hand, the emitted light gives straightforward experimental access to the polariton field. Due to the finite lifetime of cavity photons, which usually does not exceed 100 ps, a continuous decay rate of polaritons makes this system inherently out-of-equilibrium\cite{col_coh,pol_cond}. An external laser source is needed to balance this net loss rate of polaritons and replenish the microcavity.

We will focus on coherently pumped polaritons systems, in which the laser excitation energy is close to the polariton resonance energy, in contrast to non-coherently pumped systems, where the laser energy strongly exceeds this energy.  In this pumping regime no free phase is present because, through the coherent excitation, the fluid's phase is locked directly to the phase of the incoming laser field. 
Thanks to their coherence, which is inherited from the excitation laser, the resonantly excited polaritons can be described by a single wave function whose dynamics is governed by a generalized Gross-Pitaevskii equation\cite{col_coh,pol_cond,iac_review}.

Although we are dealing with non-equilibrium systems, one can still investigate the superfluid properties. One of the central aspects of superfluidity is the frictionless flow past an obstacle. In the case of equilibrium Bose-Einstein condensates, this question was addressed by Astrakharchik and Pitaevskii \cite{grisha}. Its generalization to nonequilibrium polariton quantum fluids was initiated by Carusotto and Ciuti \cite{iac_superfluid} and succesfully verified experimentally by Amo {\em et al.} \cite{amo_superfluid}. The main conclusion of these works was that the interactions between polaritons allow for a flow with strongly reduced scattering off defects below a critical velocity that coincides with the Landau critical velocity for weak defects \cite{iac_superfluid,amo_superfluid} and is lower for a strong defect \cite{amo_soliton,simon_soliton}. Above the critical velocity, Cerenkov type sound waves are observed in the weak defect case and vortex emission and soliton formation was seen in the strong defect regime.

In all the above mentioned theoretical analyses and experimental observations, the polariton nonequilibrium quantum fluids behaved largely analogously to their equilibrium counterparts. The nonequilibrium nature of the polariton quantum fluids however allows for an increased flexibility. At equilibrium, the frequency of the Bose field is set by the chemical potential \cite{lpss}, which depends uniquely on the density, whereas under coherent excitation it is set by the excitation laser frequency. The independent tunability of the density and the optical excitation frequency allows for the exploration of novel regimes in parameter space. It has been illustrated that the pump detuning $\Delta$, the energy difference between incident laser beam and the polariton resonance, allows for a new gamma of scattering physics \cite{andrei_drag,iac_zebra}. The reason is that the linear excitation spectrum, as is found in equilibrium, transforms in a range of qualitatively different spectra, which influence the resulting scattering physics profoundly.

We will show in this article the previously overlooked fact that, under certain conditions, a nonequilibrium quantum fluid can exert a  negative drag force on a defect, i.e. a force directed opposite to the flow direction. This implies the remarkable fact that freely moving defects do not remain at rest with respect to the fluid. Instead, an equilibrium velocity $v_0>0$ is reached that satisfies $F(v<v_0)<0$ and $F(v>v_0)>0$. The physical mechanism behind the negative drag force is the selective parametric amplification of the scattered waves. Both the freely tunable frequency and the polariton-polariton interactions are essential for the occurrence of this phenomenon, showing that it can only take place in nonequilibrium interacting quantum fluids.

Our paper is structured as follows: In Sec. \ref{sec:spectra} we recall the Gross-Pitaevskii equation and illustrate the corresponding excitation spectra. These results will be used to derive an expression for the drag force in Sec. \ref{sec:drag} where numerical calculations are shown for different parameter regimes. A closer look will be taken at the conditions under which the drag force can become negative in Sec. \ref{sec:dragneg}. The region in parameter space will be determined, as well as the value of the equilibrium velocities $v_0$. 

\section{The Excitation Spectrum}
\label{sec:spectra}
The time-evolution of a resonantly driven polariton fluid in a microcavity is given by a generalized Gross-Pitaevskii equation \cite{iac_review}. Only the occupation of the lower-polariton (LP) field at small momenta $\vec{k}$ is assumed, such that the LP dispersion is approximately quadratic. The description of polariton field $\Psi(\vec{r},t)$ then reduces to: 
\begin{equation}
\label{eq:GP}
\begin{split}
 \partial_t \Psi(\vec{r},t) = \Bigg[-\frac{1}{2m}\nabla^2 + g\Big|\Psi(\vec{r},t)\Big|^2-\frac{i}{2}\gamma +V(\vec{r})\Bigg]\Psi(\vec{r},t) \\ + \mathcal{F}(\vec{r},t).
\end{split}
 \end{equation}
An effective polariton mass $m$, interaction constant $g$ and a polariton decay rate $\gamma$, corresponding to a finite polariton lifetime $\tau=1/\gamma$, have been introduced. 
The pump field is taken to be a plane wave with an amplitude that is constant in space and time:
\begin{equation}
\label{eq:pump}
 \mathcal{F}(\vec{r},t)=f_p e^{i\brac{\vec{k_p}\cdot\vec{r}-\omega_p t}}
\end{equation}
In order to study scattering effects, a localized defect potential with strength $g_V$ positioned at $\vec{r}=0$ is introduced:
\begin{equation}
\label{eq:defect}
 V(\vec{r})=g_V\delta(\vec{r}).
\end{equation}
The defect can be created by an additional laser with a beam radius much smaller than the pumping laser, that locally blueshifts the exciton energy \cite{amo_lightpot}.  Alternatively, it can be formed by a variation in the cavity thickness, acting as a potential on the photonic component \cite{amo_superfluid,nardin_vort}. 

We will assume the defect to be weak, in order to apply linear response theory to obtain an approximate solution (for effects beyond linear response, see Ref. \onlinecite{emiliano}). In this treatment, a solution is proposed that consists of a mean-field steady-state solution $\Psi_{SS}$, which inherits the phase of the laser pump, and a small first-order perturbation $\delta\psi$, representing the system's response to the defect: $\Psi(\vec{r},t)=e^{i\brac{\vec{k}_p\cdot \vec{r}-\omega_p t}}[\psi_{0}+\delta\psi(\vec{r})]$.
Hence, this so-called Bogoliubov ansatz is expressed in terms of its Fourier components as \cite{iac_superfluid,andrei_drag}:
\begin{equation}
\label{eq:bog}
\delta \psi(\vec{r}) =  \frac{1}{V}
\sum_{\vec{k}}{\delta \psi (\vec{k}) e^{i \vec{k}\cdot\vec{r}}}
 \end{equation}
Substituting (\ref{eq:bog}) in (\ref{eq:GP}), along with the defect potential (\ref{eq:defect}) and the plane-wave pump field (\ref{eq:pump}) yields a coupled system of equations in $\vec{k}$-space.

 When gathering terms of first order in $g_V$ and $\delta\psi$ oscillating with the same phase, one derives an equation of the form:
 \begin{equation}
 \label{eq:lin}
 \hat{ \mathcal{L}}(\vec{k})\vectr{\delta\psi(\vec{k})}{\delta\psi^{\ast}(-\vec{k})}=\vectr{-g_V\psi_0}{g_V\psi_0^{\ast}}
 \end{equation}
The Bogoliubov operator $\hat{\mathcal{L}}$ is defined as:
 \begin{equation}
 \label{eq:eigeq}
  \hat{ \mathcal{L}}(\vec{k})\equiv \matrx{\epsilon(\vec{k})+\mathcal{E}+\vec{k}\cdot\vec{v}
  -i\frac{\gamma}{2}}{g\psi_0^2}{-g{\psi_0^{\ast}}^2}{-\epsilon(\vec{k})-\mathcal{E}+\vec{k}\cdot\vec{v}
  -i\frac{\gamma}{2}}
 \end{equation}
\begin{figure}
\centering
\includegraphics[width=\columnwidth]{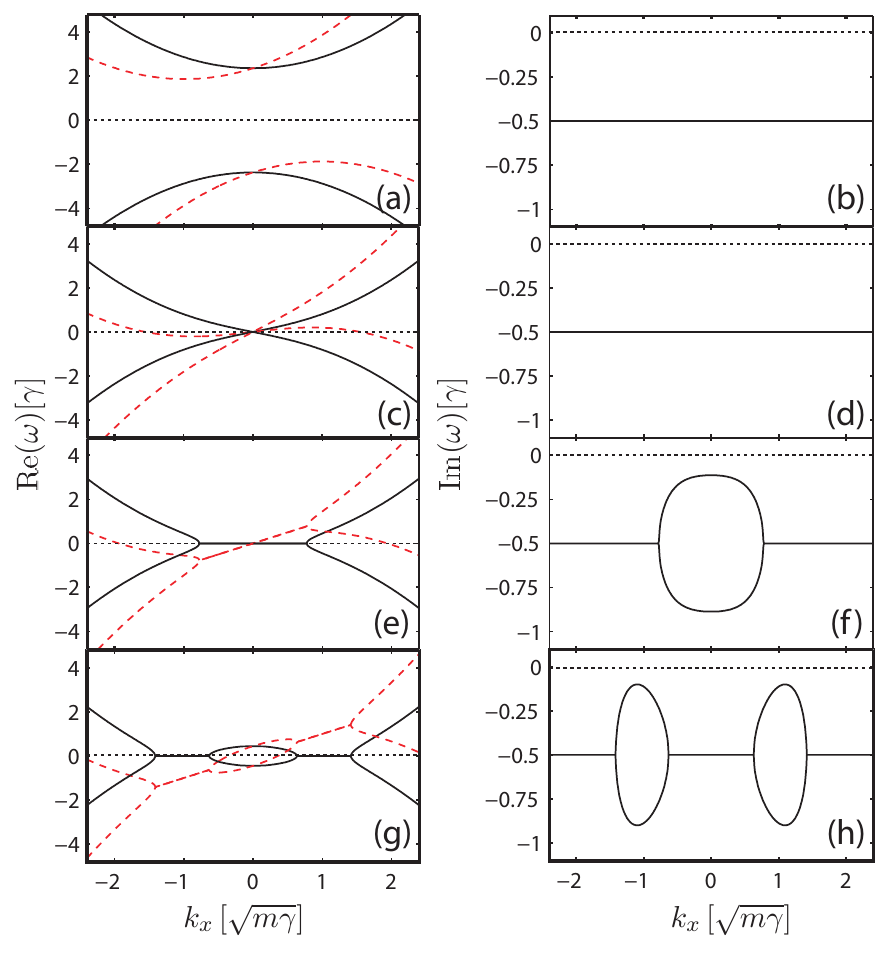}
\caption{The different spectra of a single-spin polariton fluid with $\mathcal{E}=0.3\gamma$. The left panel is the real part of the spectrum and the right the imaginary. The spectra are given for pumping detuning $\Delta=-2\gamma$ (a,b), $\Delta=0$ (c,d), $\Delta=0.3\gamma$ (e,f) and $\delta=\gamma$ (g,h).}
\label{fig:spectra}
\end{figure}
 The spectrum of excitations is then given by the eigenvalues of $\hat{\mathcal{L}}$:
 \begin{equation}
\label{eq:dispE}
 \omega_\pm(\vec{k})= \vec{k}\cdot\vec{v}-\frac{i}{2}\gamma \pm \sqrt{\epsilon(\vec{k})\brac{\epsilon(\vec{k})+2\mathcal{E}}},
\end{equation}
where we have defined
\begin{eqnarray}
 \epsilon(\vec{k})&=&\frac{k^2}{2m}-\Delta,\\
 \Delta&=&\omega_p-\brac{\frac{k_p^2}{2m}+\mathcal{E}},\\
 \mathcal{E}&=&g\norm{\psi_0}^2
\end{eqnarray}
The parameter $\Delta$ represents the laser detuning, the  energy difference between the laser pump energy and the chemical potential of the polariton fluid. 
$\mathcal{E}$ is the interaction energy. From (\ref{eq:lin}), the Bogoliubov wave-functions in $\vec{k}$-space are found to equal
\begin{equation}
\label{eq:wave}
 \delta \psi(\vec{k})=g_V\psi_0\;\dfrac{\epsilon(\vec{k})-\vec{k}\cdot\vec{v}+i\gamma/2}{\omega_+(\vec{k})\;\omega_-(\vec{k})}
\end{equation}
 As a consequence of the tunable parameter $\Delta$, the argument of the root in (\ref{eq:dispE}) is not necessarily positive. This gives rise to so-called diffusive-like spectra, in which the argument of the square root argument is negative for certain momentum modes. In general four different types of spectra can be distinguished. Examples of each type are shown in Fig. \ref{fig:spectra}:
\begin{itemize}
 \item $\Delta<0$: The argument of the root is positive for every value of $\vec{k}$. The dispersion relation of the Bogoliubov excitations reduces to the quadratic gapped dispersion for massive particles. For small $\vec{k}$, we then find that the negative detuning generates an effective excitation gap.
 \item $\Delta=0$: With zero detuning, the polariton system becomes identical to an equilibrium Bose-Einstein condensate, except for the particle decay rate $\gamma$, which comes as a global shift of the imaginary part of the excitation frequency. The real part of the Bogoliubov spectrum is then found to be linear and the excitations are massless.
 \item $0<\Delta<2\mathcal{E}$: In a circle with radius $\sqrt{2m\Delta}$ around $\vec{k}=0$ the root has a negative argument. More technically speaking, when the parameter $\Delta$ passes through zero, a bifurcation occurs in the excitation energy spectrum. As a result the spectral modes with $k<\sqrt{2m\Delta}$ have purely imaginary energy values for $v=0$. As the maximum of the imaginary part lies at $\vec{k}=0$, i.e. for a fluctuation at the laser momentum. This signals the onset of an instability of the optical bistability type \cite{my_opo,bistab_baas}.
 \item $\Delta>2\mathcal{E}$: When increasing the detuning further, a circle around $\vec{k}=0$ becomes real-valued again and the bifurcation is shifted to a ring between the radii $k=\sqrt{2m\Delta}$ and $k=\sqrt{2m(\Delta-2\mathcal{E})}$. This is the regime of parametric amplification\cite{pol_opa,my_opo}, where polaritons from the pump beam are scattered by the process $2\vec{k}_p \rightarrow \vec{k}_s+\vec{k}_i$ ($s$ and $i$ referring to signal and idler respectively. Here, the signal and idler lie around the two maxima of the imaginary part of the excitation spectrum ).
\end{itemize}

Care has to be taken for the validity of the linearization of the equations of motion around the homogeneous solution. This procedure is only valid when the excitation energies \eqref{eq:dispE} both have a negative imaginary part. In the last two cases (panels (c) and (d) in Fig. \ref{fig:spectra}), the square root contributes to the imaginary part and the requirement of linear stability puts an upper bound on the interaction energy $2\mathcal{E} < \gamma$. When this condition is violated, bistability and/or parametric instability takes place and our linearized equations of motion \eqref{eq:lin} break down.

Below, we will show that it is precisely in the cases with nontrivial imaginary part of the dispersion \eqref{eq:dispE} that the peculiar situation of a negative drag force can take place. Notice that the stability condition does not limit the physical range of the detuning $\Delta$. Thus, in principle, every value of the experimentally tunable $\Delta$ can correspond to a stable polariton system.

\section{The Drag Force}
\label{sec:drag}
The force that the flowing fluid exerts on the defect is given by\cite{grisha,andrei_drag}:
\begin{eqnarray}
\label{eq:dforce}
\vec{F}&=& \int{ d\vec{r} \norm{\Psi(\vec{r})}^2 \vec{\nabla} V(\vec{r})} 
\label{eq:defF}\\
&=&-g_V\nabla\norm{\Psi}^2\Big|_{\vec{r}=0} \label{eq:Fgradn},
\end{eqnarray}
where we have used partial integration and used the defect potential (\ref{eq:defect}). Substitution of the density profile (\ref{eq:bog}) yields:
\begin{eqnarray}
 \label{eq:imdrag}
  \vec{F}&\equiv&2g_V \int{ \frac{d\vec{k}}{(2\pi)^2} \vec{k}\; \Im \bigg[ \psi_0^{
  \ast} \delta \psi(\vec{k})   \bigg] } \\   \label{eq:fd_psi}
  &=& 2 g_V^2\norm{\psi_0}^2 \int{ \frac{d\vec{k}}{(2\pi)^2}\vec{k}\;\Im\brac{\dfrac{\;\epsilon(\vec{k})}{\omega_+(\vec{k})\omega_-(\vec{k})}}}
 \end{eqnarray}
or more explicitely when we direct the flow along the $x$-axis and transform to polar coordinates:
\begin{widetext}
\begin{equation}
\label{eq:F_expl}
F=\dfrac{2g_V^2\norm{\psi_0}^2\gamma v}{(2\pi)^2} 
\int_{0}^{2\pi} d\theta  \int_{0}^{\infty}
{ dk\;
\frac{ k^3 \; \epsilon(k)\cos(\theta)}
{\Big[\epsilon(k)\Big(\epsilon(k)+2\mathcal{E}\Big)-\big[v k \cos(\theta)\big]^2+\frac{\gamma^2}{4}\Big]^2+\Big[\gamma vk \cos(\theta) \Big]^2}}.
\end{equation}
\end{widetext}
This integral can be evaluated straightforwardly numerically. 

\begin{figure}
 \centering
 \includegraphics[width=\columnwidth]{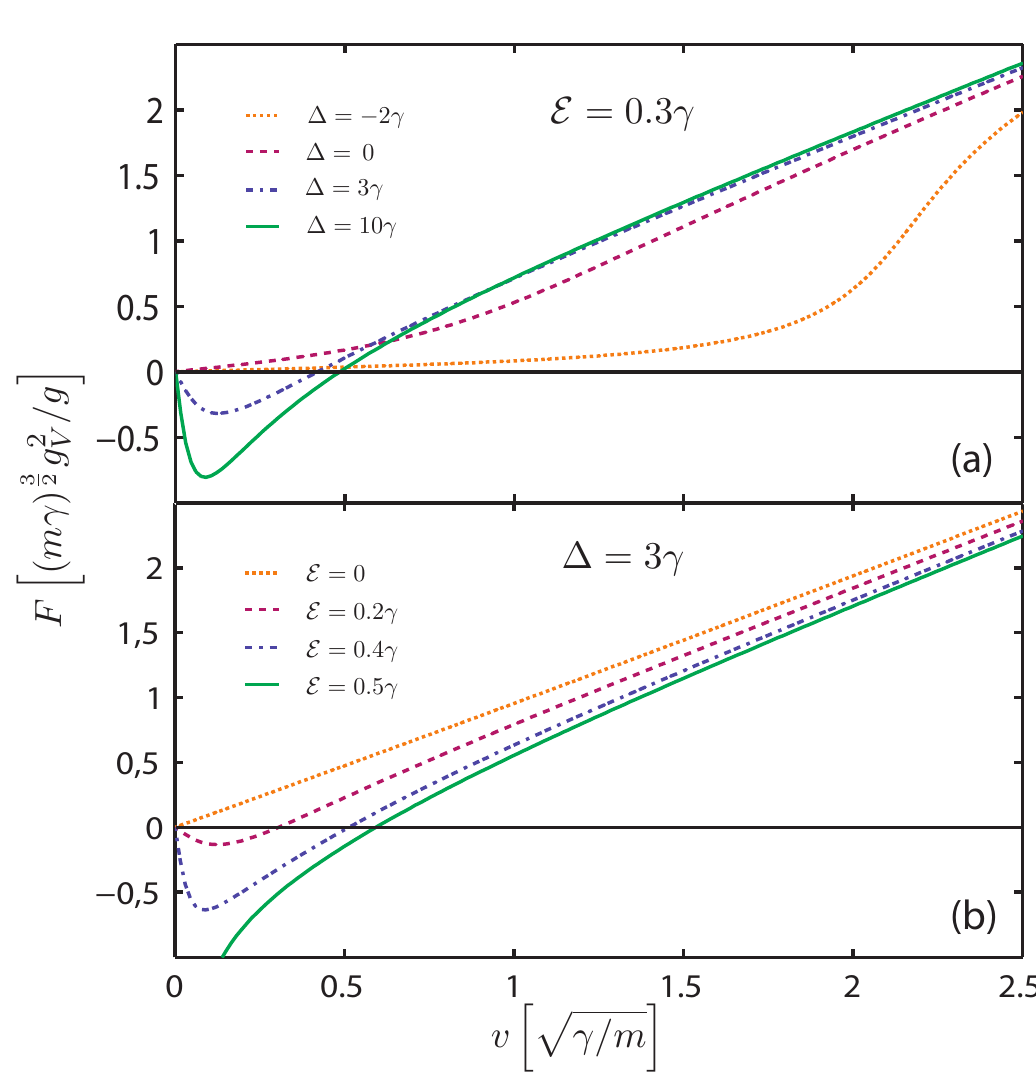}
 \caption{Panel (a) illustrates the influence of the detuning $\Delta$ on the drag force, with $\mathcal{E}=0.3\gamma$ fixed. Negative $\Delta$ induce a jump in the drag curve, whereas positive $\Delta$ can cause the drag to go below zero. Panel (b) shows the effect of interactions, with $\Delta=3\gamma$ fixed. For $\mathcal{E}=0$ only a positive drag force occurs. $\mathcal{E}=\gamma/2$ leads to a divergence of $F$ for $v\rightarrow0$. }
 \label{fig:dragcurves}
\end{figure}

In Fig. \ref{fig:dragcurves}, the influence of the pump detuning $\Delta$ and the interaction energy $\mathcal{E}$ on the drag force is illustrated by evaluating expression (\ref{eq:F_expl}). For $\Delta<0$ the quasiparticle spectrum is gapped. In the limit $\gamma \rightarrow0 $, which is a valid limit in this case, this causes a jump at a critical velocity\cite{andrei_drag} $v_c > c_s = \sqrt{\mathcal{E}/m}$. As a consequence, the drag force does not start increasing linearly at $v=v_c$, but rather goes through an abrupt jump \cite{andrei_drag}. For $\Delta>0$, it is seen on figure \ref{fig:dragcurves} that the drag force curve can become negative for small fluid velocities. At large fluid velocities, the usual positive drag force is always recovered.

\section{The negative drag force}
\label{sec:dragneg}

The observation of a negative-valued drag force in Fig. \ref{fig:dragcurves} is counter-intuitive and needs physical clarification. In Fig. \ref{fig:dragcurves}, it can be seen that a negative drag force only occurs in the diffusive or parametric amplification regimes. The appearance of a negative-valued drag force is actually a consequence of parametric scattering triggered by the presence of the defect. The selective parametric amplification of the scattered waves that have the largest imaginary part (see Fig. \ref{fig:spectra} (f,h))  results in an increased scattering in the direction of the condensate flow. This implies that the force exerted by the fluid on the is in the direction opposite to the flow. This is in contrast to the equilibrium case (superfluid or normal), where more particles are scattered backwards than forwards so that a positive drag force is obtained.

\begin{figure}
 \centering
 \includegraphics[width=\columnwidth]{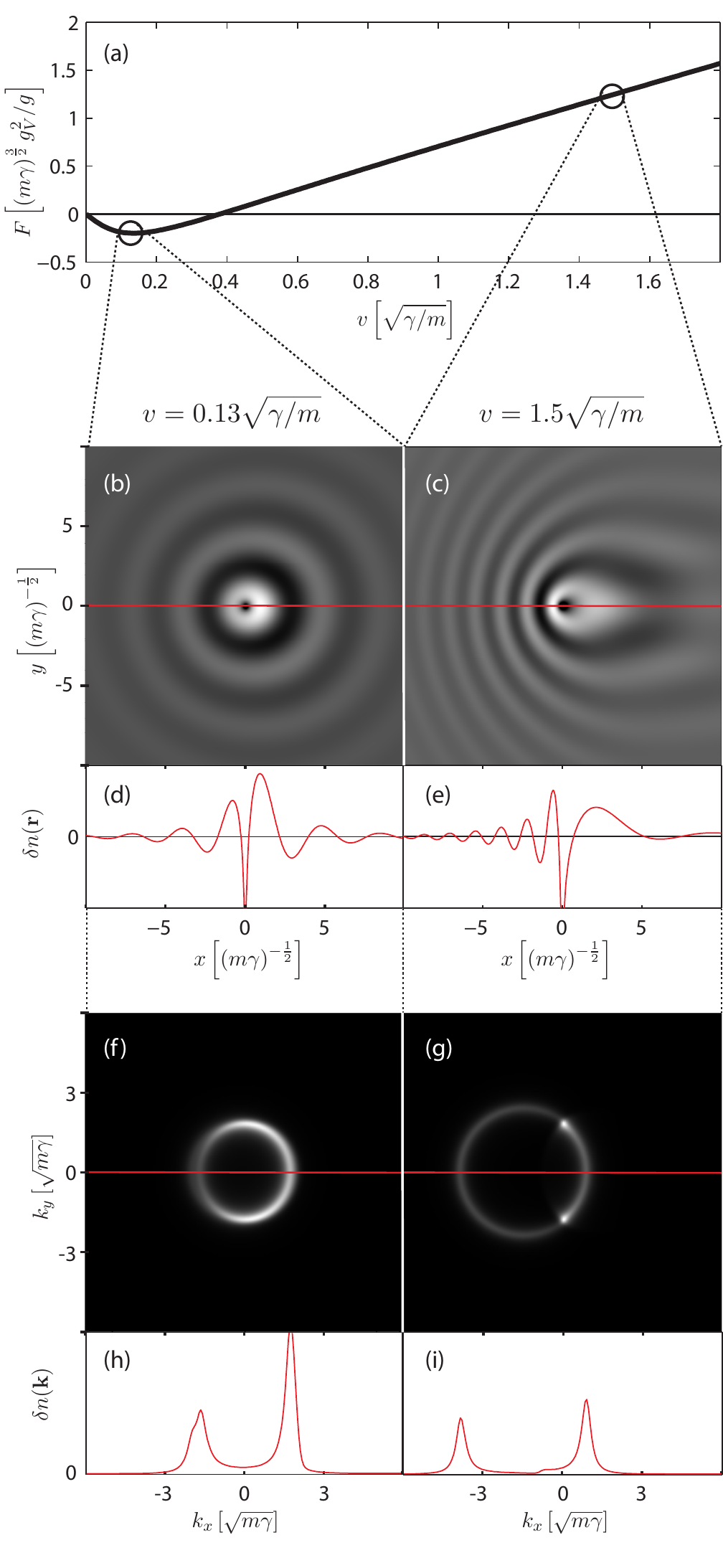}
 \caption{Panel (a) shows the drag force for a polariton condensate with $\Delta=2\gamma$ and $\mathcal{E}=0.3\gamma$. For both a sub- (left) and a supercritical velocity (right) the wave profile (b,c) with a section for $y=0$ (d,e) and the momentum distribution (f,g) with a section for $k_y=0$ (h,i) are shown. The fluid flow is oriented from left to right along the $x$-axis. The density $\delta n$ and the response function $\delta\psi(\vec{k})$ are plotted in arbitrary units. For small velocities, forward scattering is dominant, which can be deduced from the big peak for positive $k_x$ in (h). This is the origin of the drag force oriented in the direction of the condensate flow. On panel (i) we see that the peaks have moved towards negative $k_x$ for higher velocities, causing a drag force in the direction of the fluid flow the fluid flow.}
 \label{fig:profiles}
\end{figure}

This mechanism is also visible in the numerically computed momentum distribution, shown in Fig. \ref{fig:profiles}. Panel (f) shows the momentum distribution in the case of the negative drag force (see panel (a)). The cut along the $x$-axis in panel (h) clearly shows that the forward scattering peak carries more particles as compared to the backward scattering peak. 

The relation between the drag force and the momentum distribution can be made more precise by considering the photon momentum balance. When the system is in a steady state, the momentum distribution of the polaritons inside the microcavity is stationary. Photons are however constantly added by the excitation laser and reflected off and transmitted through the microcavity. The force exerted on the defect is then given by
\begin{equation}
\vec{F} =  \dot{\vec{P}}_{\rm las}-\dot{\vec{P}}_{\rm refl}-\dot{\vec{P}}_{\rm trans},
\label{eq:mombal}
\end{equation}
where $\dot{\vec{P}}_{\rm las},\dot{\vec{P}}_{\rm refl},\dot{\vec{P}}_{\rm trans}$ are the momentum per unit time carried by the excitation laser, reflected and transmitted beams respectively. 
In the Bogoliubov approximation, the momentum balance between incident, reflected and transmitted light at $\vec{k}=\vec{k}_p$ is not disturbed. Hence, the momentum transferred to the microcavity is given by the momentum carried by the scattered waves that leak out of the microcavity, yielding
\begin{equation}
\vec{F} = \gamma \int \frac{d \vec{k}}{(2\pi)^2} \; \vec{k} \;  \delta n(\vec{k}).
\label{eq:fd_mom}
\end{equation}
 We have checked that with $\delta n(\vec{k})=|\delta \psi(\vec{k})|^2$ and using Eq. \eqref{eq:wave}, the expression \eqref{eq:F_expl} is recovered.  
For the experimental measurement of the drag force, the relation \eqref{eq:mombal} and its perturbative limit \eqref{eq:fd_mom} have the advantage with respect to the defining expression \eqref{eq:defF} that it is neither necessary to measure the defect potential nor the real space polariton density with high spatial resolution. 

The physical picture that relates the negative drag force to parametric amplification is also visible in the real space density profile of the fluid. Panel (d) of Fig. \ref{fig:profiles}, corresponding to a negative drag force, show that waves are emitted in the wake of the defect. The enhanced emission in forward direction results from the selective parametric amplification of the scattered waves. When the flow velocity is increased, the wave pattern changes character from cylindrical to the `zebra cerenkov' pattern, first discussed in Ref. \onlinecite{iac_zebra}. It is the consequence of the interference of the two dominant peaks in the momentum distribution (see panel (g)). In this regime, we see again the usual pileup of density before the defect (see Fig. \ref{fig:profiles} panels (c,e)), resulting in a positive drag force. In the momentum space, we see that the Rayleigh ring shifts towards momentum states  directed opposite to the fluid flow. The average momentum of the scattered waves then lies opposite to the superflow, resulting in the familiar backward scattering of excitations and thus yielding a positive drag force. 

\begin{figure}
 \centering
 \includegraphics[width=\columnwidth]{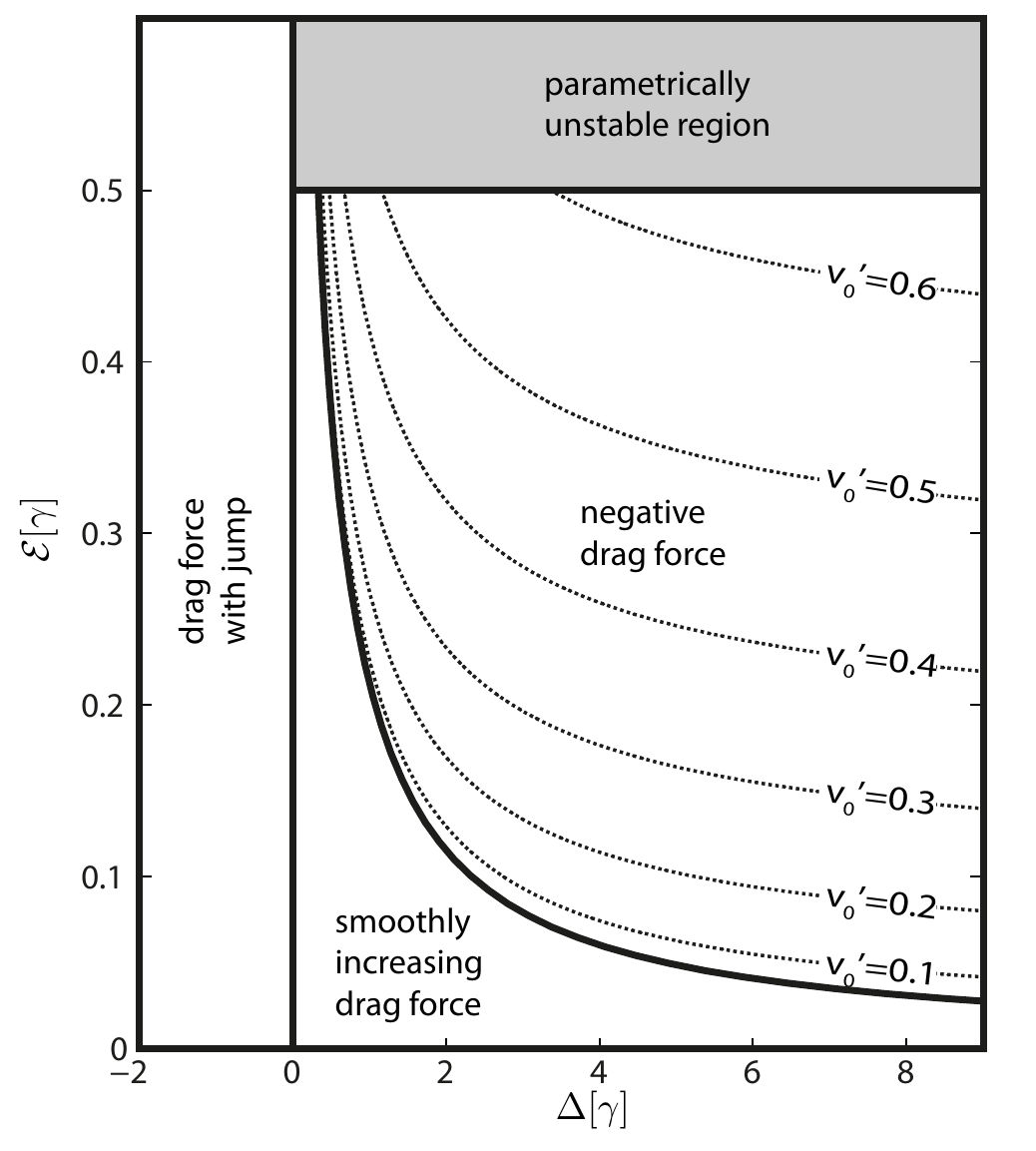}
 \caption{A phase diagram of the different types of drag-force curves that can occur in a polariton system. The unphysical region is due to $\mathcal{E}>\gamma/2$ in polariton fluids with diffusive-like spectra. The contour that separates the negative drag force from a positive one is given by (\ref{eq:bound}). The contours of the non-zero equilibrium velocities $v_0$ are shown in this region. We made use of the reduced velocity $v'_0=v_0\sqrt{m/\gamma}$. }
 \label{fig:phaseFd}
\end{figure}

Let us now turn to the systematic study of the conditions under which the negative drag force can appear in terms of the detuning $\Delta$ and interaction energy $\mathcal{E}$. The phase diagram in Fig. \ref{fig:phaseFd} indicates the different regimes of the drag force. It shows that the negative drag force exists only for sufficiently large interaction energy and detuning. In the appendix, we derive an analytic formula for the condition under which a negative drag force exists. It takes the compact form:
\begin{equation}
\label{eq:bound}
\dfrac{2\mathcal{E}\sqrt{\gamma^2-4\mathcal{E}^2}}{4\Delta\mathcal{E}-\gamma^2} <\arctan\brac{\frac{2(\mathcal{E}-\Delta)}{\sqrt{\gamma^2-4\mathcal{E}^2}}}-\frac{\pi}{2}
\end{equation}
In the limit $\Delta/\gamma\rightarrow +\infty$ this reduces to:
\begin{equation}
 \mathcal{E}>\frac{\gamma^2}{4\Delta}.
 \label{eq:cond_an}
\end{equation}
This condition may wrongly suggest that it is easier to reach the negative drag force regime at high detuning. However, the polariton density and hence the interaction energy decreases as $\mathcal{E} \propto |f_p|^2 \Delta^{-2}$ at large detuning. To meet condition \eqref{eq:cond_an}, a higher laser intensity $|f_p|^2$ is therefore required when $\Delta$ is increased.

\begin{figure}
 \centering
 \includegraphics[ width=\columnwidth]{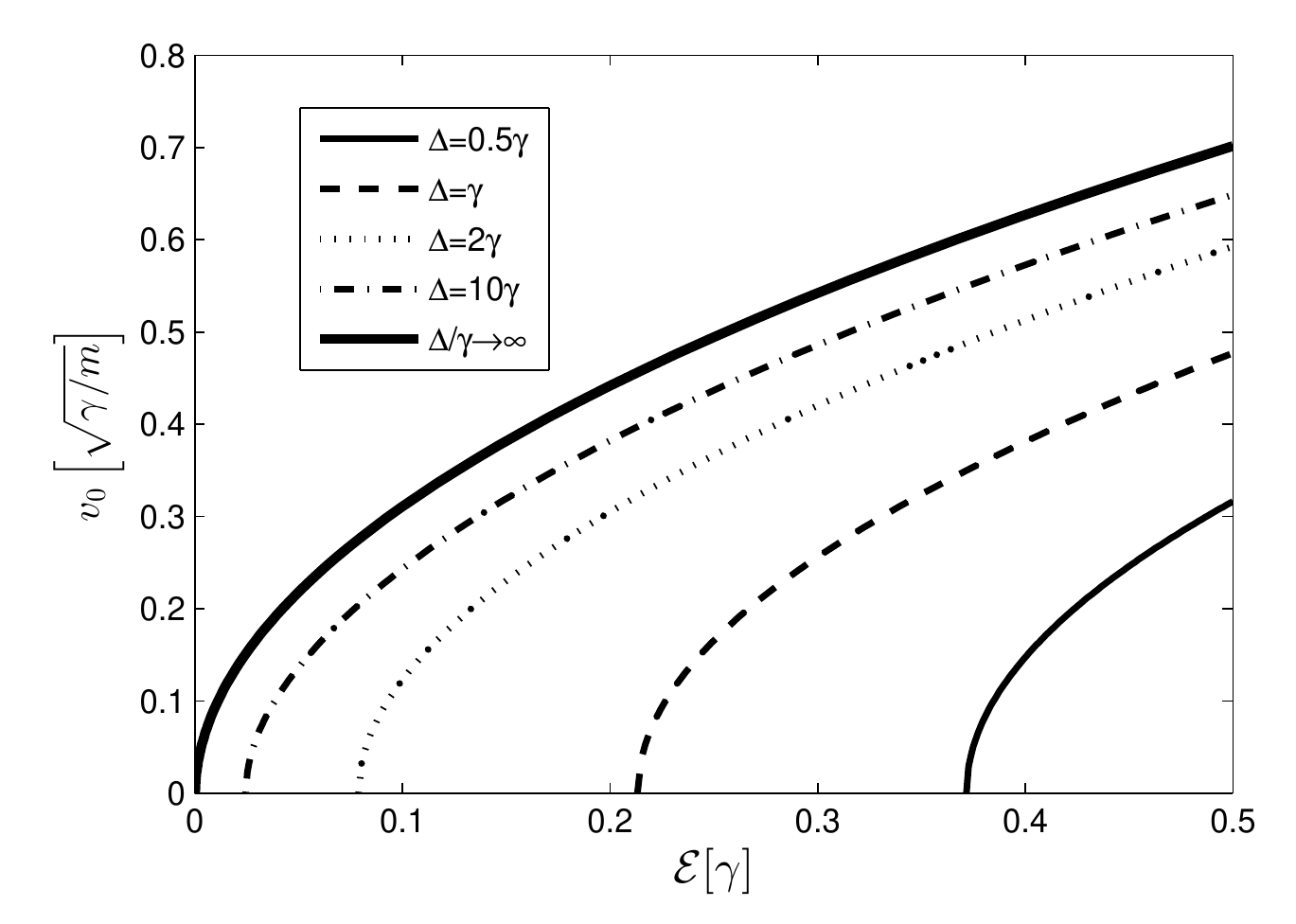}
 \caption{The non-zero equilibrium velocity $v_0$ as a function of the interaction energy $\mathcal{E}=g\norm{\psi_0}^2$ for different values of the detuning $\Delta$. In the limit $\Delta/\gamma\rightarrow \infty$, this curve converges to $v_0\approx \sqrt{\mathcal{E}/m}$.}
 \label{fig:velCurves}
\end{figure}

In Fig. \ref{fig:phaseFd}, the dashed lines show contours of the velocity $v_0$ where the drag force vanishes, as a function of detuning and interaction energy. 
The dependence of the equilibrium velocity on the bulk interaction energy is illustrated on Fig. \ref{fig:velCurves}. Numerically we found that the behaviour of the equilibrium velocity for $\Delta/\gamma\rightarrow \infty$ goes as $v_0\approx\sqrt{\mathcal{E}/m}=c_s$, which coincides with the speed of sound.

\begin{figure}
 \centering
 \includegraphics[width=\columnwidth]{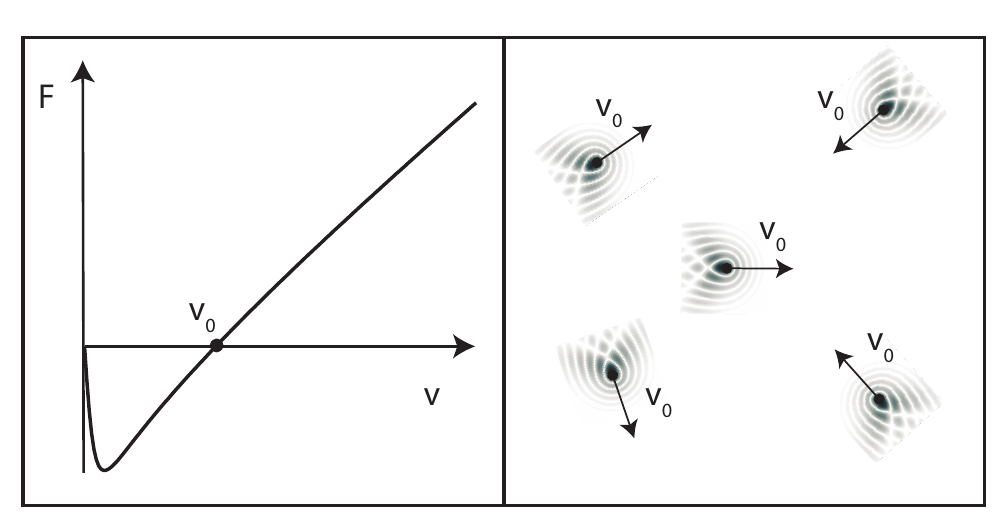}
 \caption{Mobile defects in a polariton condensate would start propagating at the equilibrium velocity $v_0$, the non-zero root of the drag force curve. }
 \label{fig:movingparticles}
\end{figure}

The velocity $v_0$ has a remarkable physical meaning when considering defects that are mobile rather than fixed with respect to the microcavity. In the regime of negative drag, the derivative $\partial F/\partial v$ is negative in the origin(see the full lines in Fig. \ref{fig:dragcurves}), implying that a small fluctuation in the velocity of the defect with respect to the fluid is amplified rather than damped. The acceleration of the defect continues until the second root of $F(v)$ is reached, at finite speed $v_0$. At that speed, the derivative $\partial F(v=v_0)/\partial v$ becomes again positive, leading to stabilization of the speed. This leads to the phenomenon that mobile defects do not remain at rest with respect to the fluid, but start to move at the speed $v_0$, in random directions (see Fig. \ref{fig:movingparticles}). The nonequilibrium situation ensures that this is not in contradiction with energy conservation: due to the excitation with a detuning $\Delta$, an ``excess energy'' is available that can be converted into kinetic energy of the impurities.

\section{Conclusions}

In conclusion, we have performed a systematic study of the behavior of the drag force as a function of interaction energy and detuning. A remarkable regime with negative drag force was uncovered. We have related this phenomenon to parametric amplification in the nonequilibrium system. An analytic condition for the existence of a negative drag force was derived. The shape of the drag force versus velocity leads tot he prediction that mobile defects will start to move at a finite speed with respect to the non-equilibrium polariton quantum fluid. The relation of the force to the momentum distribution allows for its straightforward experimental determination.

\section{Acknowledgements}
We acknowledge stimulating discussions with Francesca Marchetti, Andrei Berceanu and Iacopo Carusotto. This work was financially supported by the FWO Odysseus program and the UA-BOF.

\section*{Appendix: Calculation of the Boundary Contour}
The boundary contour given in (\ref{eq:bound}) can be calculated analytically via expression (\ref{eq:F_expl}). We want to find a root $v_0$ for $F(v)$ in the limit $v_0\rightarrow 0$. Since $F(v\rightarrow 0)$ already equals zero due to trivial symmetry considerations in the angular integral, we are looking for an additional root coming from the radial integral.
Thus, we need to find solutions of:
\begin{eqnarray}
 0&=&\int_{0}^{\infty}{ dz\;\frac{ z\;\brac{\frac{z}{2m}-\Delta}}{\bigg[\brac{\frac{z}{2m}-\Delta}\Big(\frac{z}{2m}-\Delta+2\mathcal{E}\Big)+\frac{\gamma^2}{4}\bigg]^2}}\nonumber \\
\Longleftrightarrow 0&=&\int_{0}^{\infty}{ dz'\;\frac{ z'\;\brac{\frac{z'}{2}-\Delta'}}{\bigg[\brac{\frac{z'}{2}-\Delta'}\Big(\frac{z'}{2}-\Delta'+2\mathcal{E'}\Big)+\frac{1}{4}\bigg]^2}}\nonumber 
\end{eqnarray}
We have made a rescaling of the variables: $z'=z/(m\gamma)$, $\Delta'=\Delta/\gamma$ and $\mathcal{E}'=\mathcal{E}/\gamma$ to obtain dimensionless quantities. To avoid an overload of notation, we omit the primes in the following.
Shifting the integration variable $z\rightarrow z+\Delta$ yields:
\begin{eqnarray}
 0&=&\int_{-\Delta}^{\infty}{ dz\;\frac{z^2-\Delta^2}{\Big[\brac{z-\Delta}\big(z-\Delta+4\mathcal{E}\big)+1\Big]^2}}\nonumber \\
 &=&\int_{-\Delta}^{\infty}{ dz\;\frac{z^2-\Delta^2}{\Big[ \alpha z^2+ \beta \Delta^2-2\Delta z+4\mathcal{E}\brac{z-\Delta} \Big]^2}}\;\Bigg\rvert_{\alpha=\beta=1}
\end{eqnarray}
With the use of this suggestive notation, we can write the equation as:
\begin{equation}
 \int_{-\Delta}^{\infty}{dz\left[\pd{}{\alpha}U(z,\alpha,\beta)-\pd{}{\beta}U(z,\alpha,\beta)\right]}_{\alpha=\beta=1}=0
\end{equation}
With:
\begin{equation}
 U(z,\alpha,\beta)=\frac{1}{ \alpha z^2+ \beta \Delta^2-2\Delta z+4\mathcal{E}\brac{z-\Delta}}
\end{equation}
Since we can change the order of integration and partial differentiation, the problem reduces to evaluating:
\begin{equation}
 \int_{-\Delta}^{\infty}{U(z,\alpha,\beta)\;dz},
\end{equation}
This integral of the form $\int{\brac{uz^2+vz+w}^{-1}dz}$ can be evaluated in closed form. After substituting and differentiation, one obtains expression (\ref{eq:bound}).

\end{document}